\documentclass[12pt]{iopart}
\usepackage{graphicx}

\begin{document}

\title[On the Calculation of the Inverse Isotope Effect in PdH(D)]{On the Calculation of the Inverse Isotope Effect in PdH(D): A Migdal-Eliashberg Theory Approach}

\author{S. Villa-Cort\'es and R. Baquero}

\address{Physics Department, Cinvestav-IPN\\
 Av. IPN 2508 GAM, 07360 Ciudad de M\'exico, M\'exico}

\date{\today}

\vspace{10pt}
\begin{indented}
\item[]September 2017
\end{indented}
\begin{abstract}
Replacement of hydrogen by deuterium in palladium hydride results in higher superconducting temperatures and an anomalous isotope effect that has not been yet fully explained. In this work, we try a new approach to the explanation of the inverse isotope effect in PdH(D). Our approach introduces two new aspects. First, we took into account the experimental evidence that at temperatures below 50 K, the crystal structure of PdH and of PdD is zincblende. Second, we take into account not only the influence of  the electron-phonon interaction but also the electron-electron interaction contribution to the isotope coeffiecient due tothe replacement of  deuterium in the place of hydrogen.  We used the Migdal-Eliashberg theory to perform our ab initio calculations.  We found in this picture that the electron-electron interaction is considerably reduced by the isotope substitution and is the most important factor to explain the inverse isotope effect. We found $\Delta T_c^{total}=2.224\:K$ and $\alpha= -0.3134$ in excellent agreement with the values found experimentally.
\end{abstract}


\vspace{2pc}


\maketitle


\section{INTRODUCTION}

Historically the isotope effect has been instrumental in understanding the mechanism responsible for Cooper pair formation in conventional superconductors. It gives the response of the vibration spectrum, the electron-phonon coupling and the Coulomb electron-electron repulsion to an isotopic mass change and therefore gives information on how the dynamics of the ions are involved in the value of $T_{c}$. If only phonons are taken into account the Bardeen-Cooper-Schrieffer theory (BCS) \cite{PhysRev.108.1175} predicts that the transition temperature of a single- element-superconductor goes as $T_c\propto M^{-\alpha}$ where $M$ is the isotope mass and $\alpha=0.5$. If we take into account the contribution of the Coulomb electron-electron repulsion some slight deviations from this value can occur in simple metals \cite{LEAVENS19741329}. There is no general behavior for the isotope effect coefficient. In $MgB_{2}$ \cite{MgB2} and the fullerides \cite{PhysRevLett.83.404}, for example,  the isotope coefficient is substantially reduced from the BCS value. In PdH, the compound interest in this work, the isotope coefficient is large and negative \cite{CHEN1989485}; under pressure it diminishes steadily \cite{PhysRevB.39.4110}. This is in contrast to $H_{3}S$  between 140 GPa and  180 GPa the isotope coefficient goes down very quickly and then it goes further down but more slowly  \cite{0953-2048-30-4-045011,italiano}.
An interesting behavior is found in $^{6}Li$. This element exhibits an unusually large isotope effect below 21 GPa. Further,  between 21 and 26 GPa, the superconducting isotope effect becomes inverse\cite{Schaeffer06012015}.
In general, a deviation from $\alpha=0.5$ can be a fingerprint of a non-conventional mechanism (e.g., spin fluctuations or correlated superconductivity), of anharmonicity or of an effect on the electron-electron interaction due to the isotope substitution \cite{LEAVENS19741329,PhysRevLett.111.177002}.

Here we consider the hydride system with a very unusual isotope effect, the PdH(D) system. Quite an amount of theoretical and experimental work \cite{PhysRevB.14.3630,YUSSOUFF1995549,CHEN1989485,BROWN197599,CRESPI1992427,PhysRevLett.57.2955,PhysRevLett.35.110,PhysRevB.17.141,0022-3719-7-15-015,PhysRevB.39.4110,Kara,PhysRevB.45.12405,PhysRevB.29.4140,PhysRevLett.34.144,PhysRevB.12.117,PhysRevLett.111.177002} has been done since the discovery of the inverse isotope effect in PdH(D) $(\alpha\approx-0.3)$ without reaching to a satisfactory explanation of this phenomenon. Most of the mechanism that have been proposed to explain it attribute the inverse isotope effect to  vibrational effects of Hydrogen and Deuterium, such as anharmonicity \cite{Kara,PhysRevB.45.12405, PhysRevLett.111.177002} or to  the zero-point motion \cite{PhysRevB.29.4140}, both of which have an effect the electron-phonon coupling. The contribution of the electron-electron interaction has not been yet fully considered.

Another aspect of this problem is associated to the crystal structure considered for PdH(D). Most of the work done so far considers the rocksalt crystalline structure where the hydrogen atoms are located on the octahedral sites of the fcc lattice of Palladium \cite{PhysRevB.14.3630,YUSSOUFF1995549,CHEN1989485,BROWN197599,CRESPI1992427,PhysRevLett.57.2955,PhysRevLett.35.110,PhysRevB.17.141,0022-3719-7-15-015,PhysRevB.39.4110,Kara,PhysRevB.45.12405,PhysRevB.29.4140,PhysRevLett.34.144,PhysRevB.12.117,PhysRevLett.111.177002}. Neutron diffraction techniques have been employed to study the hydrogen-atom configuration in a single-phase sample of beta-PdH at several selected temperatures. The suggested low-temperature ($T\ll55$ K) structure of this compound is one which conforms to the space group $R\bar{3}m$. This means that, depending on temperature ($T\ll55$ K), the hydrogen atoms move from their octahedral positions towards tetrahedral ones forming the zincblende structure  \cite{PhysRev.137.A483,CAPUTO-ALI,PhysRevB.78.014104,doi:10.1063/1.4901004}  For PdD something very similar occurs \cite{0295-5075-64-3-344}.  In a theoretical study, for pressures below 20 GPa at 0 K  the stable structure was found to be zincblende \cite{doi:10.1021/jp210780m}. 
So, following the facts just mentioned, in this paper we will consider the PdH(D) in the zincblende crystal structures that has not been yet considered in this problem.
Since the anharmonic hypothesis does not reproduce the experimental value of the isotope effect coefficient for PdH(D), we try in this work to study another possibility. We take into account besides the changes in the electron-phonon interaction,  the ones in the electron-electron interaction. We take  the vibrational modes to be harmonic. Under these hypothesis, we have found that using the Migdal-Eliashberg theory \cite{Daams1979,PhysRevB.12.905} it is possible to account for the experimental results reported for the inverse isotope effect in PdH(D). This analysis might be useful to study the isotope effect in other hydride systems as well \cite{203,PhysRevLett.114.157004,PhysRevLett.96.017006,PhysRevB.96.100502,0953-2048-30-4-045011,Szczniak201730,Flores-Livas2016,doi:10.1021/acs.jpcc.5b12009,0953-2048-28-8-085018,italiano}.

The paper is organized in the following way. The theory and the basic equations that supports our method are present in Section II. The technical details and results are presented in Section III and concluding remarks are contained in Sec. IV.

\section{THEORY AND BASIC EQUATIONS}

BCS theory gives for a compound with several atoms the following formula for the partial isotope effect coefficients $\alpha_{i}\equiv-d\ln T_{c}/d\ln M_{i}$ where $M_{i}$ is the mass of the different atoms in the compound and $T_{c}$ the critical temperature. The total isotope effect coefficient is given by the sum of the partial ones, namely $\alpha_{tot}=\sum_{i}\alpha_{i}$. According to Migdal-Eliashberg theory, the $\alpha_{i}$ coefficients contain information on both the electron-phonon interaction and the electron-electron repulsion \cite{LEAVENS19741329,PhysRevLett.111.177002}. If we take into account only the mass dependence of the electron-phonon interaction, Rainer and Culetto \cite{PhysRevB.19.2540} have shown that the isotope effect coefficient can be calculated from the formula

\begin{eqnarray}
\alpha_{e-ph}\left(\omega\right) & \equiv & R\left(\omega\right)\alpha^{2}F\left(\omega\right),
\end{eqnarray}

where $R\left(\omega\right)$ is given by 

\begin{eqnarray}
R\left(\omega\right) & = & \frac{d}{d\omega}\left[\frac{\omega}{2T_{c}}\frac{\delta T_{c}}{\delta\alpha^{2}F\left(\omega\right)}\right].
\end{eqnarray}

The Eliashberg function is defined as 

\begin{eqnarray}
\alpha^{2}F\left(\omega\right)= &  & \frac{1}{N\left(\epsilon_{F}\right)}\sum_{nm}\sum_{\vec{q}\nu}\delta\left(\omega-\omega_{\vec{q}\nu}\right)\sum_{\vec{k}}\left|g_{\vec{k}+\vec{q},\vec{k}}^{\vec{q}\nu,nm}\right|^{2}\label{eq:a2f-def}\\
 &  & \times\delta\left(\epsilon_{\vec{k}+\vec{q},m}-\epsilon_{F}\right)\delta\left(\epsilon_{\vec{k},n}-\epsilon_{F}\right),\nonumber 
\end{eqnarray}

where $g_{\vec{k}+\vec{q},\vec{k}}^{\vec{q}\nu,nm}$ are the matrix
elements of the electron-phonon interaction, $\epsilon_{\vec{k}+\vec{q},m}$
and $\epsilon_{\vec{k},n}$ are the energy of the quasi-particles
in bands $m$ and $n$ with vectors $\vec{k}+\vec{q}$ and $\vec{k}$
respectively. The functional derivative of the critical temperature
 with respect to the Eliashberg function is given by \cite{Bergmann1973}

\begin{eqnarray}
\frac{\delta T_{c}}{\delta\alpha^{2}F\left(\omega\right)} & = & -\left(\frac{\partial\rho}{\partial T}\right)_{T_{c}}^{-1}\frac{\delta\rho}{\delta\alpha^{2}F\left(\omega\right)}.\label{Dev F}
\end{eqnarray}

Now we can calculate the change in the critical temperature and in
the isotope coefficient as \cite{PhysRevB.19.2540} 

\begin{eqnarray}\label{eq 5}
\triangle\ln T_{c} & = & -\int_{0}^{\infty}d\omega\alpha^{2}F\left(\omega\right)R\left(\omega\right)\triangle\ln M,\label{eq:delta-tc-el-ph}
\end{eqnarray}

and

\begin{eqnarray}
\alpha_{el-ph} & = & \int_{0}^{\infty}d\omega\alpha\left(\omega\right).\label{eq:alfa cull}
\end{eqnarray}

The phonon spectrum in PdH(D) is neatly separated in two frequency regions (see Fig. \ref{fig1}). The Pd vibrations produce acoustic modes while the hydrogen (deuterium) modes give rise to the optical modes. So, by integrating Eq. (\ref{eq:alfa cull}) in the range of frequencies corresponding to Pd we can find the electron-phonon contribution to the isotope effect coefficient corresponding to this atom. In a similar manner we can calculate the corresponding contribution from hydrogen or deuterium.

Further, the information on the electron-electron contribution can be found from the corresponding Coulomb repulsion parameter $\mu^{*}$. This is a Coulomb effective potential that depends on the phonon frequency. It gives a measure on how retardation effects due to the electron-phonon interaction (due to the ions field) scale the bare Coulomb potential \cite{KNIGAVKO20022325}. 

In the Random phase approximation  it is given by \cite{ALLEN19831}

\begin{eqnarray}
\frac{1}{\mu^{*}} & = & \frac{1}{\mu}+\ln\left(\frac{\omega_{el}}{\omega_{ph}}\right).\label{eq:Mu}
\end{eqnarray}

Where $\mu=\left\langle V\right\rangle N\left(E_{F}\right)$ is the product of the average of the coulomb potential and the density of states at the Fermi level, $\omega_{el}$ is an electron energy scale and $\omega_{ph}$ is a phonon energy one. It is therefore evident from Eq. (\ref{eq:Mu}) that $ {\mu^{*}}$ depends on the ion mass through the phonon energy. Here we do not calculated $\mu^{*}$, we use the LMEE to fit it to the corresponding critical temperature (see below).

The isotope coefficient for the electron-electron interaction is given by \cite{LEAVENS19741329}
 
\begin{equation}
\alpha_{el-el}=-\frac{d\ln T_{c}}{d\ln M} .
\end{equation}

Here the critical temperature of the isotope is given by \cite{PhysRevB.30.5019}

\begin{equation}
T_{c}^{PdD}=T_{c}^{PdH}+\triangle T_{c}^{el-el} 
\end{equation}

where

\begin{eqnarray}
\triangle T_{c}^{el-el} & = & \frac{\partial T_{c}}{\partial\mu^{*}}\left(\mu_{PdD}^{*}-\mu_{PdH}^{*}\right),\label{eq:Delta tc mu}
\end{eqnarray}

and

\begin{eqnarray}
\frac{\partial T_{c}}{\partial\mu^{*}} & = & -\left(\frac{\partial\rho}{\partial T}\right)_{T_{c}}^{-1}\frac{\partial\rho}{\partial\mu^{*}}.
\end{eqnarray}

Now, if we take both contributions into account the total isotope effect coefficient is given by the following equation

\begin{eqnarray}\label{atot}
\alpha_{tot} & = & \alpha_{el-ph}+\alpha_{el-el},
\end{eqnarray}

and the total change in the critical temperature, $T_{c}$, is calculated from

\begin{eqnarray}
\triangle T_{c}^{total} & = & \triangle T_{c}^{el-ph}+\triangle T_{c}^{el-el}.
\end{eqnarray}

Then, according to Eqs. (\ref{eq:alfa cull} -\ref{atot}), to know $\alpha_{tot}$ it is necessary to know first $\mu^{*}$ and $\delta T_{c}/\delta\alpha^{2}F\left(\omega\right)$. We can get $\mu^{*}$ by solving the LMEE valid at $T_{c}$ ounce $\alpha^{2}F\left(\omega\right)$ is known and then we can calculate the functional derivative, $\delta T_{c}/\delta\alpha^{2}F\left(\omega\right)$, using the formalism of Bergmann \cite{Bergmann1973} and of Leavens \cite{LEAVENS19741329}. For an isotropic superconductor, the LMEE

\begin{eqnarray}
\rho\bar{\triangle}_{n} & = & \pi T\sum_{m}\left[\lambda_{nm}-\mu^{*}-\delta_{nm}\frac{\left|\tilde{\omega}_{n}\right|}{\pi T}\right]\bar{\triangle}_{m}.\label{Eli Ecua}
\end{eqnarray}

Where $\bar{\triangle}_{n}$ is given by 

\begin{eqnarray}
\bar{\triangle}_{n} & = & \frac{\left|\tilde{\omega}_{n}/\omega_{n}\right|\triangle_{n}}{\left|\tilde{\omega}_{n}\right|+\pi T\rho},
\end{eqnarray}

Here $\rho$ is the breaking parameter that becomes zero at $T_{c}$. The
frequency $\tilde{\omega}_{n}$ is 

\begin{eqnarray}
\tilde{\omega}_{n} & = & \omega_{n}+\pi T\sum_{m}\lambda_{nm}sig(\omega_{m}),
\end{eqnarray}

and $i\omega_{n}$ are the Matsubara frequencies, $i\omega_{n}=i\pi T\left(2n+1\right)$ with $n=0,\pm1,\pm2\ldots$. The coupling parameter $\lambda_{nm}$ is defined as 

\begin{eqnarray}
\lambda_{nm} & = & 2\int_{0}^{\infty}\frac{d\omega\omega\alpha^{2}F\left(\omega\right)}{\omega^{2}+\left(\omega_{n}-\omega_{m}\right)^{2}}.\label{eq:lambdas}
\end{eqnarray}

$\lambda_{nn}$ is the known electron-phonon interaction parameter.

\section{TECHNICAL DETAILS AND RESULTS}

In this section we use the analysis showed previously  to study the isotope effect of PdH with the zincblende crystal structure. Since several works consider the rocksalt crystal structure to be the proper one and attribute the inverse isotope effect to hanarmonic affects in the vibrational spectra, we also study the isotope effect taken into account this structure. Finally we compare the results from both structures.

\begin{figure}
\begin{center} 
\includegraphics[width=8.4cm]{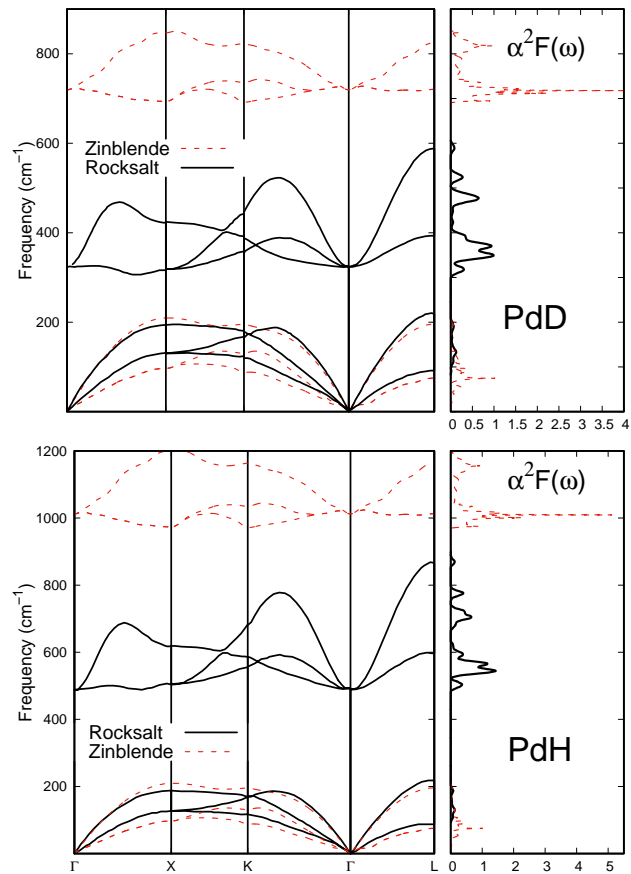}
\caption{\label{fig1}(color online) Phonon spectra and Eliashberg function calculated for PdH(D) with the zincblende crystal structure and the ones for the rocksalt crystal structure taken from reference \cite{PhysRevLett.111.177002}.}
\end{center} 
\end{figure}

For the zincblende structure (Fig. \ref{fig1}), the phonon spectra and the Eliashberg function were obtained using the Quantum - Espresso suite code \cite{0953-8984-21-39-395502}. We used the density functional perturbation theory \cite{0953-8984-21-39-395502,RevModPhys.73.515} and the scalar relativistic pseudo potentials of Pardue and Zunger (LDA) \cite{PhysRevB.23.5048}. We used 150 Ry cutoff for the plane-wave basis and a 32 X 32 X 32 mesh for the BZ integration in the unit cell. For the force constants matrix we used a 16 X 16 X 16 mesh. The sum over $\vec{k}$ in Eq. (\ref{eq:a2f-def}) required a 72 X 72 X 72 grid. For the rocksalt structure we take the data from reference \cite{PhysRevLett.111.177002}, which include anharmonic effects. 

For both crystal structures the phonon spectrum and therefore, the Eliashberg function, is divided into two frequency regions. The acoustic region, from the Pd vibrations, is in general similar for both structures and it ranges from 0 to approximately 250 $cm^{-1}$. For the two structures the optical frequency region, from the Hydrogen and Deuterium vibrations, is separated by a frequency gap from the acoustic ones. The hydrogen optical region is higher than that for the deuterium one in both structures and there is a shift to higher frequencies from the the rocksalt to the zincblende structure (see Fig. \ref{fig1}).

To solve the LMEE, we used a cutt-off frequency, $\omega_{cutoff} = 10 \omega_{ph}$ where $\omega_{ph}$  is the maximum phonon frequency, to cut the sum over the Matsubara frequencies. With the $\omega_{cutoff}$ chosen the only adjustable parameter left is $\mu^{*}$. We fitted it by solving the LMEE to get the experimental $T_c$. In Tab. \ref{tab:table1} we include the fitted values of $\mu^{*}$.

\begin{figure}
\begin{center} 
\includegraphics[width=8.4cm]{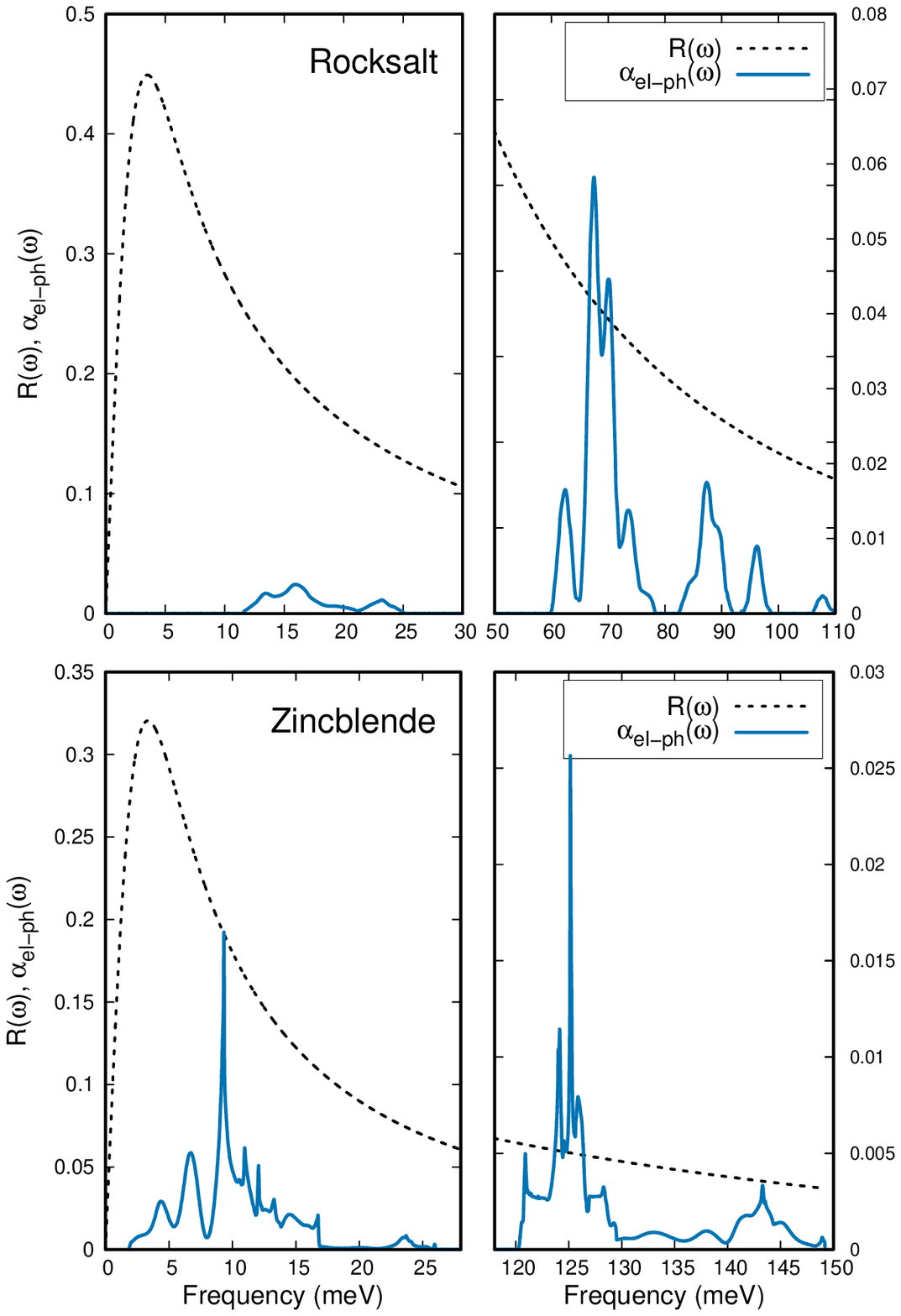}
\caption{\label{fig2}(color online) Weighting function $R\left(\omega\right)$ and differential isotope effect $\alpha_{el-ph}\left(\omega\right)$ calculated for PdH at $T_{c}=8.8\:K$ for both rocksalt (top) and zinckblende (bottom) structures. Left, acustic (Palladium) frequency region. Right, optical (Hydrogen) frequency region. }
\end{center} 
\end{figure}

\subsection{The zincblende structure}

In Fig. \ref{fig2} we show the calculated weighting function $R\left(\omega\right)$ which determine the contribution to the total isotope effect from the electron-phonon interaction around a frequency $\omega$ due to the substitution of hydrogen by deuterium. First we analyze the case of the system with zincblende structure. Although $R\left(\omega\right)$ is very small, it does not become negative on the optical frequency region. This means that the superconducting state does not change with the isotope mass replacement if only this interaction is taken into account. Now, integrating $\alpha_{e-ph}\left(\omega\right)$ over this region (Eq. \ref{eq:alfa cull}), we calculated  the isotope effect coefficient contribution to be $+0.0556$ from the electron-phonon interaction. The electron-phonon  contribution to the change in the critical temperature, $\Delta T_c^{e-ph}$, can be computed from Eq. (\ref{eq 5}). We find that  $\Delta T_c^{e-ph} = -0.332 \:K $ whereas the contribution due to the electron-electron interaction (Eq. \ref{eq:Delta tc mu}),  $\Delta T_c^{e-e} = +2.556\:K$. It is obvious that the electron-phonon contribution is almost negligible as compared to the electron-electron one. Further, we calculated  the isotope effect coefficient contribution to be $-0.369$ from the electron-electron interaction. This gives a total isotope coefficient $\alpha^{total}= -0.3134$ in remarkable agreement to the experimental value (see Tab. \ref{tab:table1}). Furthermore, this theory predicts a critical temperature for PdD 

 \begin{eqnarray}
T_c^{PdD}= T_c^{PdH}+ \Delta T_c^{total}=11.024\:K,
\end{eqnarray}

this critical temperature for the isotope PdD is very close to the experimental value of 11.05 K \cite{PhysRevB.39.4110}. We found therefore that for this system the enhancement of the critical temperature is mainly due to the change in the Coulomb parameter, Eq. (\ref{eq:Delta tc mu}).

\begin{table}[h]
\caption{\label{tab:table1}Calculated $\mu^{*}$, $\Delta T_{c}$ and $\alpha$ values for PdH(D) with the rocksalt and zincblende structures. The $\mu^{*}$ values where fitted to get the correct experimental critical temperature by solving the LMEE. $T_{c}$ values are given in K. Experimental \cite{PhysRevB.39.4110} and theoretical \cite{PhysRevLett.111.177002} values of $T_{c}$ and $\alpha$ are presented as well.}
\footnotesize
\begin{tabular}{cccccc}
\br
 & Zincblende & Rocksalt & Expt. \cite{PhysRevB.39.4110}& \multicolumn{2}{c}{From \cite{PhysRevLett.111.177002}} \tabularnewline
\br
 & $\mu^{*}$ & $\mu^{*}$& $T_{c}$&$\mu^{*}$ & $T_{c}$\tabularnewline
\tabularnewline
PdH & 0.062677 & 0.035795 & 8.8 & 0.085 & 5\tabularnewline
PdD & 0.022094 & 0.028629 & 11.05 & 0.085 & 6.5\tabularnewline
\br
\tabularnewline
$\alpha_{el-ph}$ & 0.05556 & 0.36448 &  &  & \tabularnewline
$\alpha_{el-el}$ & -0.36841 & -0.13254 &  &  & \tabularnewline
\textbf{$\alpha^{total}$}& \textbf{-0.3129} & \textbf{0.2319} & \textbf{-0.32889} & \multicolumn{2}{c}{\textbf{-0.38}} \tabularnewline
\tabularnewline
\mr
\tabularnewline
$\Delta T_c^{el-ph}$ & -0.332 & -1.962 &  &  & \tabularnewline
$\Delta T_c^{el-el}$ & 2.556 & 0.846 &  &  & \tabularnewline
\textbf{$\Delta T_c^{Total}$} & \textbf{2.2246} & \textbf{-1.1168} & \textbf{2.25} & \multicolumn{2}{c}{\textbf{1.5}} \tabularnewline
\br
\end{tabular}

\end{table}

\subsection{The rocksalt structure}

Even though, as we mention above, this is not the structure that prevails at temperatures $T\ll55$ K, we now proceed to study the system in the rocksalt structure since several works consider it in their study of this problem. In this case $R\left(\omega\right)$ is not so small as compared to the zincblende system, and remains positive in all the optical frequency region. By integrating $\alpha_{e-ph}\left(\omega\right)$ over this region we get the isotope effect coefficient contribution to be $+0.36448$ from the electron-phonon interaction. This means that the superconducting state is highly affected by the isotope mass substitution if only this interaction is taken into account. For the contribution to the corresponding  change in the critical temperature in this case,  we find  $\Delta T_c^{e-ph} = -1.962 \:K $. Now if we calculate the corresponding change due to the electron-electron interaction, we find $\Delta T_c^{el-el} = 0.846\:K$. This gives a total change in the critical temperature $\Delta T_c^{Total}=-1.1168$. Further, we calculated  the isotope effect coefficient contribution to be $-0.13254$ from the electron-electron interaction. This gives a total isotope coefficient $\alpha^{total}= 0.2319$. Then, for this system the electron-phonon contribution to the superconducting state is highly modified by isotope mass substitution whereas the electron-electron one is less important as is expected in a conventional superconductor, however the values of $\Delta T_c^{Total}$ and $\alpha^{total}$ does not match the experimental ones.

Our results for the rocksalt crystal structure contrast with the ones from  reference \cite{PhysRevLett.111.177002}, there they found the value of the total isotope coefficient to be $\alpha=-0.38$. This difference can be explained because we use a different approach.  Although both analyzes were carried out within the framework of the Migdal-Eliashberg theory, we consider how the changes in the electron-phonon and the electron-electron interactions due by the isotope substitution contribute to the total isotope effect. In reference \cite{PhysRevLett.111.177002} they consider the Coulomb parameter $\mu^{*}$ to be constant and got the total isotope coefficient form $\alpha=-\frac{d\ln T_{c}}{d\ln M}$, however their calculated critical temperature does not match the experimental ones (see Tab. \ref{tab:table1}). Therefore, both approaches fails to explain the inverse isotope effect for PdH(D) when the rocksalt crystalline structure is considered the proper one for this system. 

\section{CONCLUDING REMARKS}

In conclusion, we have found that if we consider the zincblende crystal structure as the proper one for PdH(D) at  temperatures below ($T\ll55$ K) in agreement with experiment, the inverse isotope effect can be explained by taking the electron-electron contribution into account. We found that the electron-phonon contribution is much less important. We reproduced the experimental measured values for the isotope coefficient and the change in the critical temperature. Our work provides a simple and direct explanation of the observed inverse isotope effect in PdH. We have also considered the rocksalt crystal structure for PdH(D) and were unable to reproduce the experimental results under this assumption. 

\ack
The authors acknowledge to the general coordination of information and communications technologies (CGSTIC) at CINVESTAV-IPN for providing HPC resources on the Hybrid Cluster Supercomputer "Xiuhcoatl", that have contributed to the research results reported within this paper. S. Villa-Cort\'es acknowledges the support of Conacyt-M\'exico through a PhD scholarship.

\section*{References}
\bibliographystyle{unsrt}
\bibliography{EII_IOP}

\end{document}